%% file: cikm2023paper.tex
\documentclass[sigconf]{acmart}

\AtBeginDocument{%
  }

\usepackage{indentfirst}
\usepackage{hyperref}
\begin{document}

\title{Learning Similarity among Users for Personalized Session-Based Recommendation from hierarchical structure of User-Session-Item}

\author{Jisoo Cha}
\email{jisoo.cha@yonsei.ac.kr}
\affiliation{%
  \institution{Yonsei University}
  \city{Seoul}
  \country{South Korea}
}

\author{Haemin Jeong}
\email{hmjung@yonsei.ac.kr}
\affiliation{%
  \institution{Yonsei University}
  \city{Seoul}
  \country{South Korea}
}

\author{Wooju Kim}
\email{wkim@yonsei.ac.kr}
\affiliation{%
  \institution{Yonsei University}
  \city{Seoul}
  \country{South Korea}
}


\begin{abstract}
  \input{000abstract}
\end{abstract}




\maketitle
\pagestyle{plain}

\section{Introduction}
The recommendation system is a system that identifies the user's interest and recommends appropriate items. In particular, the session-based recommendation system(SRS) takes into account the chronologically-ordered time sequences of interaction between user and items to capture the intent of users that changes from time to time. However, it is difficult to properly model sessions with traditional SRS, because of the higher-order correlation of item transitions. Also, traditional SRS deal sessions as anonymous, that causes a problem that the user's demographic information or available user information such as identification number disappears.

To address this problem, several models have been proposed. A traditional model of SRS is a MC-based method\cite{sahoo2012hidden,rendle2010factorizing,shani2005mdp}. The hidden Markov model using MC-based methods models the user's intention probabilistic and uses it for recommendation. Recently, methodologies using neural networks have also been proposed for recommendation systems to consider the increasing amount of data and limited learning time. Recurrent Neural Network(RNN)-based methodologies have been widely used in SRS systems because of their capability for modeling sequential data\cite{hidasi2015session,li2017neural,tang2018personalized}. 

However, since session-based recommendation systems using RNN are struggle to model the correlation between complex item transitions, a session-based recommendation system using graph structure has been proposed. The session-based recommendation system using the graph structure recommends the next item by making the items in a session as a node and its chronological connectivity an edge\cite{wu2019session}. At this time, the information of the item updated by Graph Neural Network(GNN) layers is pooled through certain mechanism, to create an representation of the session graph and recommendations are performed based on similarity to the entire items. This is semantically equivalent to the graph classification problem. 

However, the recommendation system that has emerged so far has the problem of processing sessions anonymously, so that useful personal information is not employed in the model. Anonymous session-based recommendation systems are modeled on the premise that all users have the same preference for all items, so there is a problem that individual special preferences are not reflected in the modeling. To solve this problem, a personalized session-based recommendation system(PSRS) has been proposed. PSRS adds a module to learn individual information to the framework of the existing recommendation system\cite{quadrana2017personalizing,pang2022heterogeneous}. Many approaches to PSRS have used a method of modeling user or session and node as one heterogeneous graph. These graphs can update heterogeneous nodes under the structure of user-session-item by transferring information through heterogenous GNN layers. Updating the user's information to the combined information of the item and session would be the reasonable approach. In addition, to model with peronal information, it would be an appropriate approach to model using not only the current user's information but also similar user's information. These two modeling methods often considered and studied in personalized session-based recommendation systems field.

To counter the issues mentioned above, we developed a user simplicity powered session based recomender (USP-GNN). We will first start with a session graph from the session data under the hierarchical structure of user-session-item. At the same time, we will build a global heterogeneous graph using the connectivity between the items of the session and the users who own the session. If the GNN is applied to the global heterogeneous graph created in this way, nodes can be updated through information from heterogeneous nodes. Unlike previous approaches of PSRS that simply use user embedding, we developed a user embedding attention module that can obtain weighted user embeddedings with additional information from sessions. In addition, we introduce contrastive loss that distinguishes similar users by comparing user embeddedings learned by local session graph, and user embeddedings learned by global session graph. Our main contributions are as follows:

\begin{itemize}
\item We defined a global heterogeneous graph to effectively model user information in SRS tasks, and a heterogeneous GNN layers. This allows the user node to receive information from the item node and update it. Since this global heterogeneous graph is constructed prior to learning, it is effective because it affects the execution time of the model in constant time.
\item We developed a user attention module that can update user information from sessions. Existing works mainly updated user embedding with item embedding, but we built weighted user embedding using the similarity between session embedding and user embedding, to model a special preference of user's individual session.
\item We proposed contrastive learning between the local user embedding and global user embedding. Since heterogenous GNN layers with global heterogeneous graph tends to smooth individual user's embedding, introducing additional contrastive loss has the effect of alleviating this.
\end{itemize}
The rest of the paper was constructed as follows. In section 2, we briefly review previous studies related to the personalized session-based recommendation system. And in section 3, we will introduce the notations used in our model and several preliminaries. Next, section 4 will introduce our proposed model, and we will analyze performances of proposed model in comparison with baseline models in section 5. Finally, we draw conclusion in section 6.

\section{Related Works}
In this chapter, we will review the related work on Session-Based Recommendation.

\textbf{Traditional Methods.} In the early SRS, modeling was conducted based on items that appear simultaneously for multiple users, rather than directly modeling sequences. A representative methodology is matrix factorization, which has been mainly used to model fragmentary relationships between users and items\cite{koren2009matrix,mnih2007probabilistic}. However, matrix factorization models only on the presence or absence of interaction between users and items, so there is a limit to capturing the intentions of users that change from time to time. To solve this problem, a markov chain-based methodology has been proposed.\cite{sahoo2012hidden,rendle2010factorizing,shani2005mdp} The Markov chain based method is a method to model a user's sequence based on the Markov assumption, but it has the difficulties in modeling sequential patterns of complex sequences.

\textbf{Deep Learning based Methods.} Recently, deep learning-based methods have been actively proposed in session-based recommendations. Early deep learning-based methodos typically come with RNN-based methodologies and CNN-based methodologies\cite{li2017neural,liu2018stamp,hidasi2015session}. GRU4REC\cite{hidasi2015session} modeled the item interaction sequence using GRU layers. NARM\cite{li2017neural} is a session-based recommendation system with an encoder-decoder structure that utilizes GRU and attention mechanisms to derive hidden representation for sessions. STAMP\cite{liu2018stamp} has attempted to model user intentions using attention mechanisms and multi-layer perceptron.

\textbf{Graph Neural Network(GNN).} GNN is a promising field in deep learning recently, because it can effectively handle graphs, which are unstructured topologies of data. Graph Convolution Network(GCN)\cite{kipf2016semi} applied trainable filters to the graph spectrum using Chevyshev polynomials, and GraphSAGE\cite{hamilton2017inductive} proposed an effective aggregation method for inductive representation learning for graphs. In addition, Graph Attention Network(GAT)\cite{velivckovic2017graph} defined aggregation function as an attention network and it performed well in downstream tasks such as node classification.

\textbf{GNN based Methods.} Thus, a GNN-based methodology has been proposed in SRS to model the higher order correlation between items in RNN, CNN-based methodologies. SR-GNN\cite{wu2019session} is the first GNN-based methodology proposed for SRS, which applies the Gated Graph Neural Network(GGNN)\cite{li2015gated} by constructing graph from item sequences. GCE-GNN\cite{wang2020global} configured a global graph to learn item embedding in two ways: local and global, and combined them to recommend appropriate items.

\textbf{Personalized Session based recommendation.} GNN-based session embeddings have an issue of anonymizing sessions, failing to utilize all available information given the data structure of the user-session-item. Therefore, combining user information with sessions is a reasonable approach, which has recently become a hot topic in session-based recommendations. Based on the historical information of the user, HRNN\cite{quadrana2017personalizing} performed user-representation propagation on the GRU layer by combining the user's information with the session embedding. \cite{zhang2020personalized} learned the historical session of the user by creating the user's unified presentation using the attention network. HG-GNN\cite{pang2022heterogeneous} used other user's historical session with heterogeneous graph for more precise recommendation.

\section{Preliminaries}
\label{sec:preliminaries}
\input{030preliminaries}

\section{Proposed Model}
In this chapter, we will elaborate on our proposed model USP-GNN(User Similarity Powered GNN). Our model is largely composed of four parts. As you can see from the Figure 3, the first step is to construct a local session graph and a global heterogeneous graph from the input session given the user's information. Second, the information of the node is updated through the GNN layers and the information of the node is graph pooled to construct the session embeddings of each local and global graph. The third part is the step of updating user embeddings by applying attention modules to session embeddings and user embeddings of local graphs. Final part is contrastive learning part that updates user embeddings with the similarity of users.

\subsection{Build Local Session Graph and Global Heterogeneous Graph}

\textbf{Local Session Graph.} Given input session $S = [v_{t_1}, v_{t_2}, ..., v{t_n}]$, local session graph $G_l=(V_l,E_l)$ can be constructed as Figure 2. Let $V_s = {v_1, v_2, v_m}$ as user-item interactions in given session $S$, and define $E$ as item's chronological connectivity to build directed graph. It can be present as the form of adjacency matrix. Note that this graph ignores duplicated edges. Following \cite{wu2019session}, we put unique items in session as $V_s$, we can build adjacency matrix $A_s$ for session $S$ as shown in Figure 1. In other words, adjacency matrix $A$ has as many rows as the number of unique items in the session, and columns as twice. If we put $v_i$ as $i$-th item in row and $v_j$ as $j$-th item in column, $A_{i,j}$ is the connectivity of item $v_i$ and $v_j$ in adjacency matrix $A$. And normalizing constraint is as follows.

\begin{equation}
  \sum_{j=1}^{|V_s|} A_{i,j} \ if\ 0<j \le |V_s|
\end{equation}
\begin{equation}
  \sum_{j=|V_s|+1}^{2|V_s|} A_{i,j} \ if\  |V_s|<j<2|V_s|
\end{equation}

\begin{figure}[h]
  \centering
  \includegraphics[width=\linewidth]{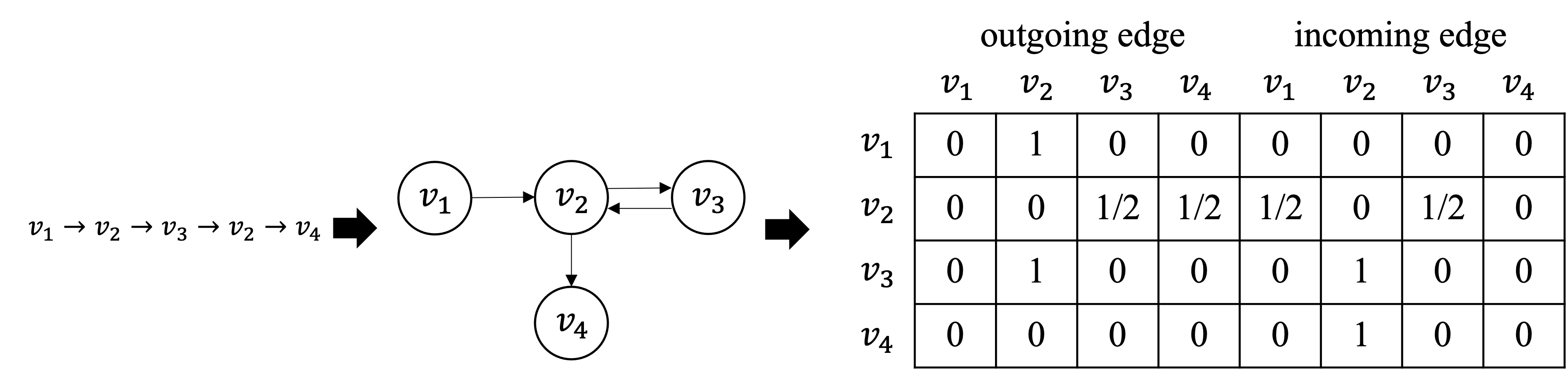}
  \caption{Example of Adjecency Matrix.}
\end{figure}

\textbf{Global Heterogeneous Graph.} Global Graph $G_g = (V_g,E_g)$ can be constructed before train step, with all sessions in train data and user index. The node set $V_g$ composed of 2 types of nodes, \textit{ItemNode} and \textit{UserNode}. Edge type can be defined by three combinations with these node type: \textit{u2i-edge} connects from \textit{UserNode} to \textit{ItemNode}, \textit{i2u-edge} connects from \textit{ItemNode} to \textit{UserNode}, and \textit{i2i-edge} connects between \textit{ItemNode}.

\begin{figure}[h]
  \centering
  \includegraphics[width=\linewidth]{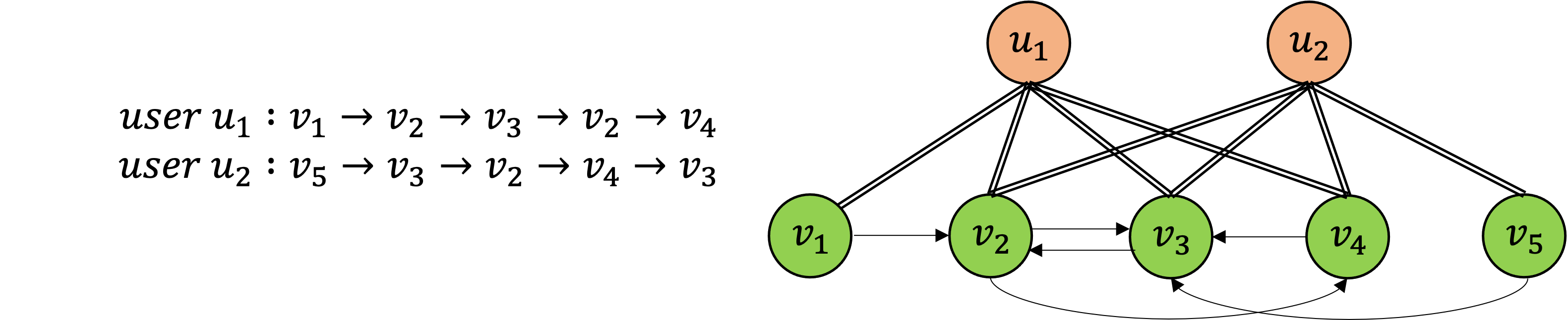}
  \caption{Example of Global Heterogeneous Graph Construction.}
\end{figure}

\textbf{Graph Neural Network(GNN).} To aggregate and update nodes in graph, we used several GNN layers for message passing function. Update function for node $v$ is as following below.

\begin{equation}
  m_v^{k+1} = \sum_{u \in N(v)} AGGREGATE_k(h_v^k, h_u^k, e_{vu})
\end{equation}
\begin{equation}
  h_v^{k+1} = UPDATE_k(m_v^{k+1}, h_v^k)
\end{equation}

where $h_v$ is the representaion for node $v$, and $k$ is order of GNN layer. $e_{vu}$ is the edge connects between node $v$ and node $u$. 
\begin{equation}
  R_G = READOUT(\{h_v^K| v \in G\})
\end{equation}

We apply different graph filters to properly model different graph types: local session graph and global heterogeneous graph. Gated Graph Neural Network(GGNN)\cite{li2015gated} is oftenly used to update session graph, since it can propagate messages with time order information, and can apply reset and update gate to filter out useless or useful information. The aggregation process for local sessiongraph is as follows.

\begin{equation}
  m_v^{k+1} = \sum_{u \in N(v)} \mathbf{W_k}h_u^k + \mathbf{b^k}
\end{equation}
\begin{equation}
  h_v^{k+1} = GRU(m_v^{k+1}, h_v^k)
\end{equation}

To apply graph neural network to update global heterogeneous graph, we apply different graph filters depending on the predefined edge types : \textit{u2i-edge}, \textit{i2u-edge}, and \textit{i2i-edge}. Aggregation process for this heterogeneous graph is like down below.

\begin{equation}
  m_v^{k+1} = \frac{1}{|N(v)|} \sum_{u \in N(v)} \mathbf{W_k} h_v^k e_t^{uv}
\end{equation}
\begin{equation}
  h_v^{k+1} = f(\mathbf{W_t}[m_v^{k+1}\mathbin\Vert h_v^{k}] + \mathbf{b_t^k})
\end{equation}

Where $t$ is the type of edge, and $e_t^{uv}$ is the indicator representing whether type $t$ edge is connected between node $u$ and $v$. Message from neighbor nodes are aggregated by linear transform layer and averaged by number of neighbor nodes. And node $v$ is updated by weight matrices defined by type of edges.

\begin{figure*}[h]
  \centering
  \includegraphics[width=\linewidth]{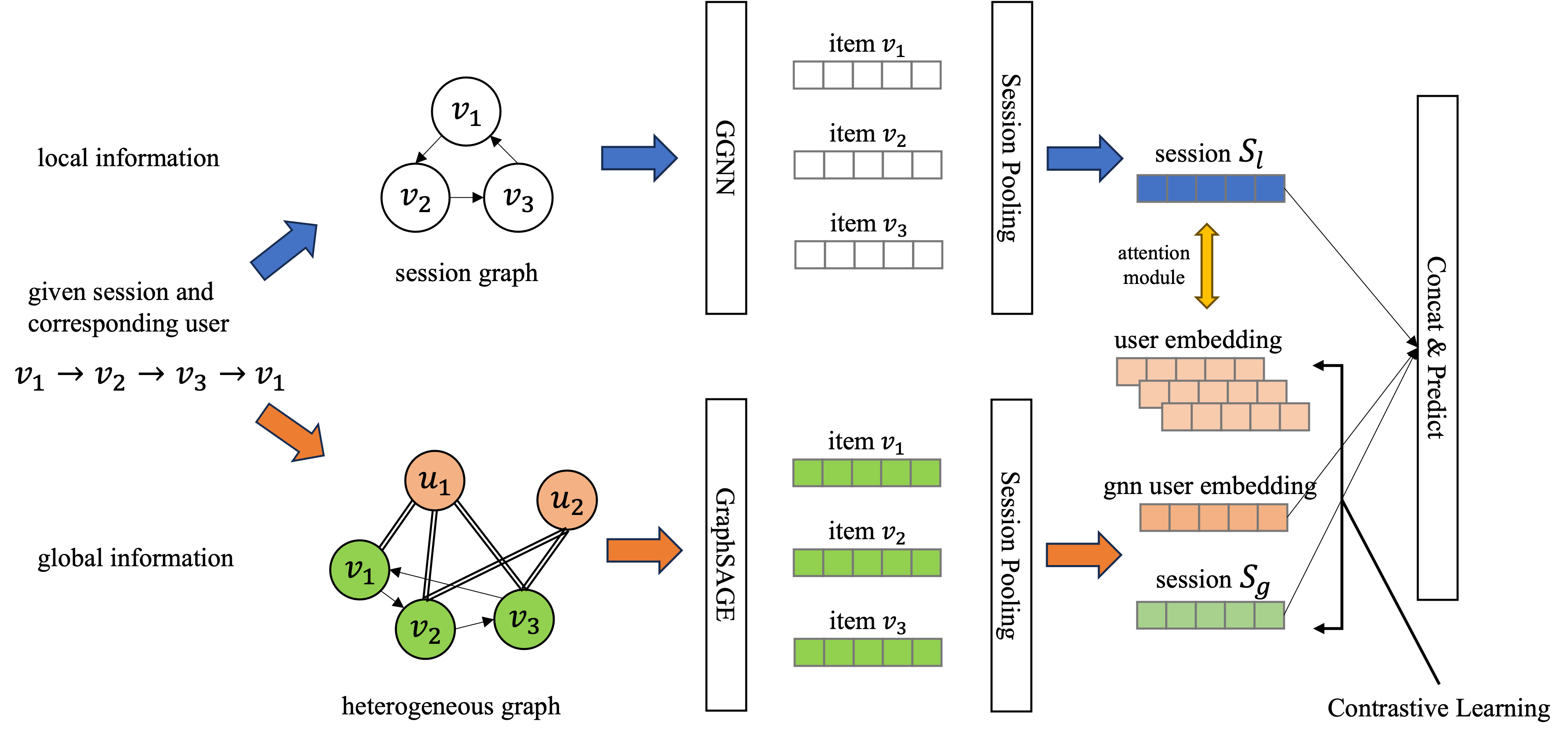}
  \caption{Our Proposed Model.}
\end{figure*}

\textbf{Construct Session Embedding.} After message propagating, we use graph pooling mechanism to obtain the whole representation of given graph $G$. Graph pooling process is as the same as readout phase. In this phase, readout function constructs representaion vector of graph from node embedding that updated by $T$ GNN layers. Basic Formula of readout function is as down below. We concatenate local embedding and global embedding and apply linear transform as $READOUT$ funcction, following same approach as \cite{wu2019session} for contruct session embedding. The equation is as follows:

\begin{equation}
  s_g = \sum_{i=1}^n \alpha_i v_i
\end{equation}
where
\begin{equation}
  \alpha_i = \mathbf{q}^\top(\mathbf{W_1}v_i + \mathbf{W_2}v_n + c)
\end{equation}
\begin{equation}
  s = \mathbf{W_3}[s_l \mathbin\Vert s_g]
\end{equation}

$s_l$ is the local preference of session $s$, suppose that user's next click will depend only current state. We set $s_l$ as chronogically last node embedding of session $s$. $s_g$ is the global preference of session $s$, obtained by computing similarity between last item of the session and the others, and applying weighted sum operation over these similarity $\alpha$ and node embedding $v_i$. As mentioned above, we concatenate there local and global preference embedding and apply linear transform layer to construct pooled embedding of graphs from node embedding.

\subsection{Combining User embeddings and Session embeddings.}

\textbf{User-Session Similarity.} To personalize SRS, properly combining user embedding known to be effective for recommendation task according to previous works. We propose \textit{UserSessionSimNet}, a user-session attention module that updates user embedding with the session embeddings of that user. For session embeddings of user, we compute similarity between each session embedding and user embedding $u_i$. And apply weighted sum by computed similarity. Formula as below:

\begin{equation}
  m_{i,j} = (\mathbf{W_q}u_i) \cdot {(\mathbf{W_k}s_j)}^\top
\end{equation}

\begin{equation}
  \alpha_{i,j} = softmax(m_{i,j}) = \frac{exp(m_{i,j})}{\sum_{\{k|s_k \in u_i\}}exp(m_{i,k})}
\end{equation}

\begin{equation}
  u_i^{updated} = \sum_{\{j|s_j \in u_i\}}\alpha_{i,j}\mathbf{W_v}u_i
\end{equation}

Linear Transformation layers that maps query, key, value embeddings to another space are denoted as $\mathbf{W_q}$, $\mathbf{W_k}$ and $\mathbf{W_v}$ respectively. Since single user $u_i$ can occupy multiple sessions, there is multiple similarity between session embedding and user embedding. So we apply softmax layer to this value by each user, for effectively compute similarity over whole batch. We than compute weighted sum based on this similarity, and update original user embedding $u_i$ to $u_i^{updated}$. This updated user embedding is combined with representaion of \textit{UserNode} updated by global heterogeneous graph.

\textbf{Final Hidden Embedding.} Final hidden embedding is consisted of session embedding from local session graph $s_{local}$ and combination of node embedding from global heterogeneous graph. To combine node embedding of two types of nodes: \textit{ItemNode} and \textit{UserNode}, we follow approaches by \cite{pang2022heterogeneous}, that compute similarity between session embedding $s_{global}$ and user embedding $u_{global}$ from global heterogeneous graph to obtain final session preference $s_{final}$.
\begin{equation}
  \beta = \sigma(\mathbf{W_s}[s_{global} \mathbin\Vert u_{global}])
\end{equation}
\begin{equation}
  s_{final} = s_{local} + (\beta \cdot s_{global} + (1-\beta) \cdot u_{global})
\end{equation}

\subsection{Contrastive Learning by User Similarity}

We propose contrastive learning for discriminating user embeddings. After propagating informations in the global heterogeneous graph by GNN, the updated representation of \textit{UserNode} can lead to decrease of model performance, since GNN layer and user embedding update module introduce perturbation to \textit{UserNode}. So we define contrastive learning\cite{chen2020simple} for this situation. This module compares original user embedding and updated user embedding, viewing updated user embedding as an data augmentation. This process makes similar user become close in embedding space, and vice versa. We let percentage of negative samples compared to positive samples as a hyperparameter. 

\subsection{Recommendation and Compute Loss} 
  
We conduct next-item recommendation task based on the final preference embedding $s_{final}$. Computing this embeddings and initial item embeddings, we can provide item recommendation and compute recommendation loss $\mathcal{L}_{recom}$ to train model. 

\begin{equation}
  \hat{y_i} = softmax(s_{final}^\top v_i^{(0)})
\end{equation}
  
\begin{equation}
  \mathcal{L}_{recom} = -\sum_{i=1}^{|V|} (y_i log(\hat{y_i}) + (1-y_i)log(1-\hat{y_i}))
\end{equation}
We combine aforementioned contrastive loss $\mathcal{L}_{cont}$with recommendation loss to make total loss $\mathcal{L}_{total}$. Also, we set the ratio of contrastive loss within total loss as a hyperparameter, called lambda.

\begin{equation}
  \mathcal{L}_{total} = (1-\lambda) \cdot \mathcal{L}_{recom} + \lambda \cdot \mathcal{L}_{cont}
\end{equation}

\section{Experiments}
In this section, we perform experimental setup and in-depth analysis to verify the performance of our model.

\subsection{Datasets}

We used two datasets for validatation. Properties of dataset summarized in Table 1.

The first dataset is the \textit{TVwatching} dataset. The data set is data from 300 viewers over a six-month period based on TV viewing records. A user interaction is defined as watching a program, an item is defined as a program, and a session is defined as a sequence of program viewing records. In raw data, sessions are randomly divided with 180 seconds interval because sessions were not previously divided. In addition, TV programs that appeared less than 5 times out of all viewing records were excluded, and data with a session length of 2 or more were used. For each user, we sort the viewing sequence in chronological order, and 80 percent of the previous session was used as a train set and 20 percent of the next session as a test set. \textit{TVwatching} dataset is private and not gonna be publicly released.

The second dataset is the \textit{AppUsage}\cite{shepard2011livelab} dataset. The \textit{AppUsage} dataset is application usage history data for the iPhone. A user interaction is defined as app use. In addition, the item becomes an app, and the session becomes an app usage record sequence. Following \cite{xie2021personalized}, system default apps removed from train or test step. Since raw data is not previously divided into sessions, sessions were divided in 600 seconds along \cite{xie2021personalized}. And we filtered out sessions with less than 3 sequence lengths. And we sort app usage sequence for each user in chronological order, and 80 percent of the previous session was used as a train set and 20 persent of the next session as a test set.

\input{table2}

\subsection{Comparisons and Parameter Settings}

We conducted a comparative experiment by selecting baseline comparison to verify the performance of our USP-GNN model. The corresponding models are as follows.

\begin{itemize}
\item GRU4REC\cite{hidasi2015session} is a typical RNN-based SRS model. It used the basic GRU module and update the parameters by calculating the top1 loss between the recommended item and the correct item.
\item NARM\cite{li2017neural} is a RNN-based recommendation system model using an additional attention module to create session embeddings.
\item SR-GNN\cite{wu2019session} is a graph-based recommendation system model that uses GNN to create session graphs, and it updates the information of items and create session embeddings.
\item GCE-GNN\cite{wang2020global} is a graph-based recommendation system model that proposes global graph to utilize global information as well as current sessions of users.
\item HRNN\cite{quadrana2017personalizing} is a personalized sequential recommendation model that executes recommendations using the user's historical session.
\item HG-GNN\cite{pang2022heterogeneous} is a personalized session-based recommendation model that implemented as constructing a user's historical session as a heterogeneous graph.
\end{itemize}

\textbf{Evaluation Metrics.} To evaluate the recommendation models and comparing between them, we use popular ranking evaluation metrics for SRS, called Hit Rate(\textit{HR@k}) and Mean Reciprocal Rank(\textit{MRR@k}), following \cite{wu2019session,wang2020global}.

\textbf{Hyperparameter Settings.} We implemented our model based on the \textit{PyG} framework\cite{fey2019fast}. We use Adam Optimizer\cite{kingma2014adam} and uses scheduler that decreases learning rate by every step size. Since SRS systems are tend to easily overfit in general, we adopt relatively small learning rate: $\{1e-4, 5e-5, 1e-5\}$. We select batch size and embedding size between $\{128,256,512\}$ by grid search. We let negative sample ratio and lambda as hyperparameters, that controls ratio of negative samples in contrastive loss and ratio of contrastive loss contributes to total loss respectively. We then search the optimal negative sample ratio and lambda by experiments.

\input{table3}

\subsection{Results}

\textbf{Comparison with Baseline Methods.} To analyze the overall performance of our proposed model, we conducted comparative experiments with other models that achieved state-of-the-art on the session-based recommendation task. The overall performance comparison is shown in Table 2. 
Our model outperformed the state-of-the-art model on a given dataset on several comparison metrics. Compared to baseline methods, the GNN-based model showed higher overall performance than the RNN family model, that means GNN-based model can be seen as more effective in reflecting user preferences in modeling. GRU4REC was not comparable in terms of performance. Nevertheless NARM showed the highest performance among the RNN family models, but experiment results shows that it has low performance overall to the comparison target model.

The GNN-based model showed good performance overall. SR-GNN is the first model to borrow a graph structure in a recommended system, so it can be seen as the baseline of the GNN-based model, which records significantly higher performance than GRU4REC, the baseline of the RNN-based model, proving that it was an appropriate approach to graph model sessions for GNN application in SRS. A high-performance model with SR-GNN as a base model is GCE-GNN, which showed the highest performance in SRS models without user information. In particular, several indicators show performance close to state-of-the-art, indicating that the global information of the item is effective in SRS.

Our model outperformed several indicators of the state-of-the-art model. Proposed model surpassed HG-GNN, current state-of-the-art model in PSRS, especially on metrics HR@3, MRR@3, which are more challenging than other metrics. In other cases, HG-GNN may perform slightly better. Our model has something in common with HG-GNN like global heterogeneous graphs, but there are differences in how information is delivered to user embedding. In addition, our model introduced additional contrastive loss to give an additional learning tasks in the model, that turned out to be effective in increasing performance.

\textbf{Ablation Analysis.} We performed ablation study on two datasets, \textit{TV Watching} and \textit{AppUsage}, to verify the efficiency and performance of the individual modules of the model we developed. The results of ablation study are shown in the Table 3. Our full model performed best on almost every metric. Based on ablation studies, the module contributing the largest performance difference was session embedding. There was a big performance difference in both \textit{HIT@k} and \textit{MRR@k} metrics. In addition, when global session embedding was excluded, there was a big difference in performance, although not as much as that of local session embedding. Finally, when excluding \textit{UserSessionSimNet}, one of our major contributions, most performance indicators showed a decline. It can also be seen that most performance indicators are lower than the standard full model when contrastive loss is not applied.

\input{table_ablation}

\textbf{Experiments on different lambda.} We measure the performance by varying Lambda's parameters, which determine the ratio of contrastive loss, to verify the performance of contrastive learning on the user embeddings we developed. Figure 4 shows the experimental results on lambda on \textit{TV Watching} dataset. Also, the Lambda experiment on \textit{AppUsage} can also be found in Figure 4. As you can see from the chart, loss constraint lambda showed balanced performance of \textit{HIT@k} and \textit{MRR@k} at around 0.3, but it tends to vary slightly depending on the data.

\begin{figure}[h]
  \centering
  \includegraphics[width=\linewidth]{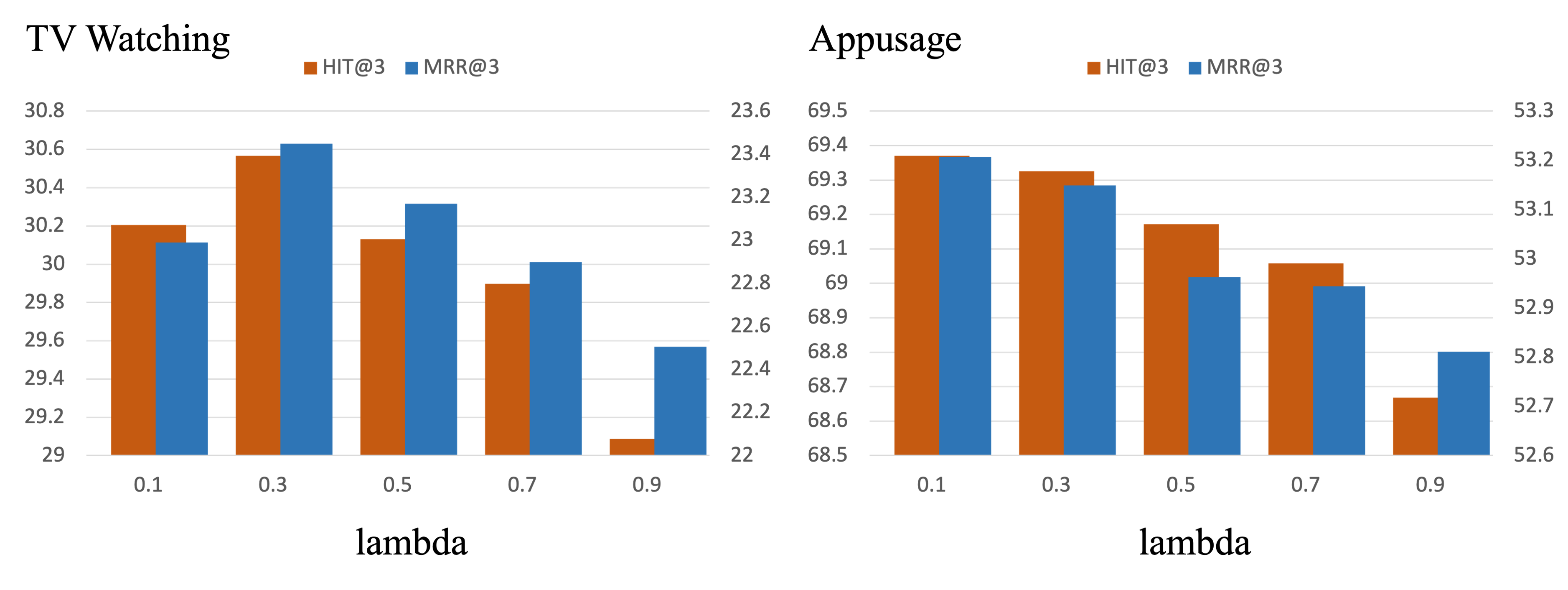}
  \caption{Experiments conducted on different lambda on two datasets.}
\end{figure}

\begin{figure}[h]
  \centering
  \includegraphics[width=\linewidth]{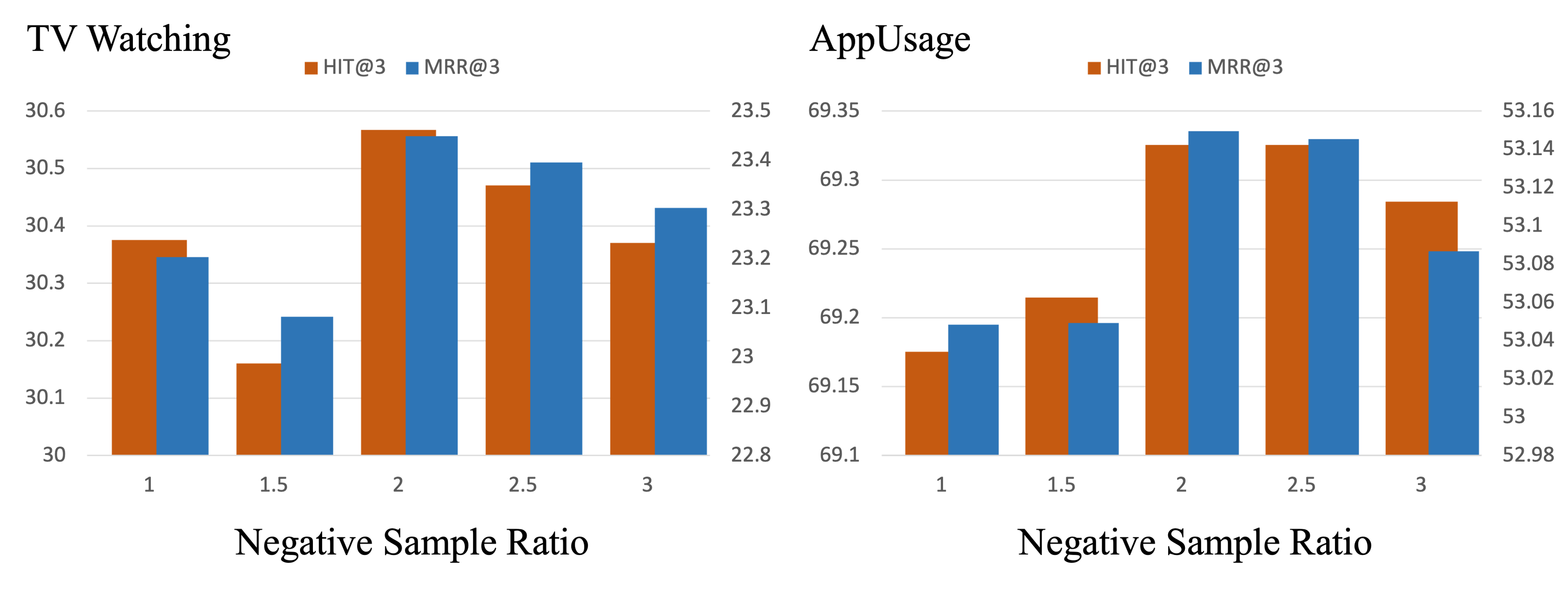}
  \caption{Experiments conducted on different negative sample ratio on two datasets.}
\end{figure}

\textbf{Experiments on different negative sample size.} We also conducted experiments by varying the negative sample ratio to verify how negative sample size affects contrastive loss. The negative sample ratio is the ratio of negative samples to positive samples, as it is called. Although it varies depending on the dataset, it showed the best performance when the negative sample ratio was 2. Too small or too large negative sample ratio is rather showed worse performance.

\section{Conclusion}
\label{sec:conclusion}
\input{060conclusions}

\section{Acknowledgements}
My work is based on a PyG implementation of SR-GNN\footnote{\url{https://github.com/lfywork/SRGNN_PyG}}.
Part of my achievement is theirs.

\bibliographystyle{ACM-Reference-Format}
\bibliography{bib_cikm}

\end{document}

%% file: 000abstract.tex
The task of the session-based recommendation is to predict the next interaction of the user based on the anonymized user's behavior pattern. And personalized version of this system is a promising research field due to its availability to deal with user information. However, there’s a problem that the user's preferences and historical sessions were not considered in the typical session-based recommendation since it concentrates only on user-item interaction. In addition, the existing personalized session-based recommendation model has a limited capability in that it only considers the preference of the current user without considering those of similar users. It means there can be the loss of information included within the hierarchical data structure of the user-session-item. To tackle with this problem, we propose USP-SBR (abbr. of User Similarity Powered - Session Based Recommender). To model global historical sessions of users, we propose \textit{UserGraph} that has two types of nodes - \textit{ItemNode} and \textit{UserNode}. We then connect the nodes with three types of edges. The first type of edges connects \textit{ItemNode} as chronological order, and the second connects \textit{ItemNode} to \textit{UserNode}, and the last connects \textit{UserNode} to \textit{ItemNode}. With these user embeddings, we propose additional contrastive loss, that makes users with similar intention be close to each other in the vector space. We apply graph neural network on these \textit{UserGraph} and update nodes. Experimental results on two real-world datasets demonstrate that our method outperforms some state-of-the-art approaches.

%% file: 030preliminaries.tex
If the personalized recommendation model data is hierarchically configured, it can be represented into hierarchichal data structure of users, sessions, and items from the upper layer. The user may have a plurality of sessions, and the session includes a plurality of items. Items and users have their own number, but sessions can always be changed because they are time-ordered arrangements of items.

\textbf{Notations.} Let $V = \{v_1, v_2, ..., v_{|V|}\}$ and $U = \{u_1,u_2,...,u_{|U|}\}$ denote an item set and user set, respectively. And user $i$'s $j$th session $S_{u_{ij}} = [v_{u_{ij},1}, v_{u_{ij},2}, ..., v_{u_{ij},n}]$ can defined following this user-item structure when the sessions are chronogically ordered. Note that session is a list that allows duplicate items. And each user can have multiple sessions $S_u = \{S_{u_{i1}},S_{u_{i2}}, ..., S_{u_{im}}\}$. Given sessions and follwing session information, the goal is to predict the next item $v_{{ij},1}$ right after the session $S_{u_{ij}}$.

%% file: table2.tex
\begin{table}
    \caption{Dataset descriptions.}
    \begin{tabular}{c|cc|}
    \cline{2-3}
                                                  & \multicolumn{1}{c|}{TV Watching} & AppUsage \\ \hline
    \multicolumn{1}{|c|}{\# of items}             & 9,441                            & 2,288    \\ \cline{1-1}
    \multicolumn{1}{|c|}{\# of users}             & 301                              & 34       \\ \cline{1-1}
    \multicolumn{1}{|c|}{\# of training sessions} & 59,182                           & 260,780  \\ \cline{1-1}
    \multicolumn{1}{|c|}{\# of test sessions}     & 14,950                           & 64,763   \\ \cline{1-1}
    \multicolumn{1}{|c|}{mean session length}     & 21.03                            & 8.48     \\ \hline
    \end{tabular}
\end{table}

%% file: table3.tex
\begin{table*}[]
\caption{Experiment results.}
\begin{tabular}{|l|rrrrrr|rrrrrr|}
\hline
\multicolumn{1}{|c|}{datasets} & \multicolumn{6}{c|}{TV Watching}                                                                                                                                           & \multicolumn{6}{c|}{AppUsage}                                                                                                                                              \\ \hline
\multicolumn{1}{|c|}{models}   & \multicolumn{1}{c|}{HR@3} & \multicolumn{1}{c|}{HR@5} & \multicolumn{1}{c|}{HR@10} & \multicolumn{1}{c|}{MRR@3} & \multicolumn{1}{c|}{MRR@5} & \multicolumn{1}{c|}{MRR@10} & \multicolumn{1}{c|}{HR@3} & \multicolumn{1}{c|}{HR@5} & \multicolumn{1}{c|}{HR@10} & \multicolumn{1}{c|}{MRR@3} & \multicolumn{1}{c|}{MRR@5} & \multicolumn{1}{c|}{MRR@10} \\ \hline
GRU4REC                        & 4.21                      & 5.98                      & 8.81                       & 2.98                       & 3.39                       & 3.76                        & 52.68                     & 54.62                     & 58.15                      & 48.09                      & 48.53                      & 49.01                       \\ \cline{1-1}
NARM                           & 30.07                     & 35.76                     & 43.39                      & 23.33                      & 24.63                      & 25.65                       & 68.04                     & 78.42                     & 85.83                      & 52.97                      & 55.37                      & 56.40                       \\ \cline{1-1}
SR-GNN                         & 16.01                     & 18.28                     & 22.24                      & 13.93                      & 14.44                      & 14.89                       & 47.63                     & 63.87                     & 74.11                      & 30.93                      & 34.68                      & 36.08                       \\ \cline{1-1}
GCE-GNN                        & 30.56                     & 36.69                     & 44.50                      & 23.00                      & 24.40                      & 25.45                       & 68.05                     & 78.32                     & 85.72                      & 52.78                      & 55.15                      & 56.17                       \\ \cline{1-1}
HRNN                           & 16.39                     & 19.57                     & 24.33                      & 12.91                      & 13.64                      & 14.27                       & 42.43                     & 44.42                     & 47.46                      & 37.05                      & 37.51                      & 37.92                       \\ \cline{1-1}
HG-GNN                         & 30.56                     & \textbf{36.81}            & \textbf{45.00}             & 22.85                      & 24.27                      & 25.37                       & 69.04                     & \textbf{79.45}            & \textbf{87.87}             & 53.09                      & \textbf{55.48}             & \textbf{56.65}              \\ \cline{1-1}
USP-GNN                        & 30.56                     & 36.63                     & 44.80                      & \textbf{23.44}             & \textbf{24.83}             & \textbf{25.92}              & \textbf{69.32}            & 79.23                     & 87.23                      & \textbf{53.14}             & 55.42                      & 56.52                       \\ \hline
\end{tabular}
\end{table*}

%% file: table_ablation.tex
\begin{table*}[]
\centering
\caption{Ablation experiment results.}
\resizebox{\textwidth}{!}{\begin{tabular}{c|rrrrrr|rrrrrr|}
\cline{2-13}
\multicolumn{1}{l|}{}                              & \multicolumn{6}{c|}{TV Watching}                                                                                                                                              & \multicolumn{6}{c|}{AppUsage}                                                                                                                                                 \\ \cline{2-13} 
\multicolumn{1}{l|}{}                              & \multicolumn{1}{c|}{HIT@3} & \multicolumn{1}{c|}{HIT@5} & \multicolumn{1}{c|}{HIT@10} & \multicolumn{1}{c|}{MRR@3} & \multicolumn{1}{c|}{MRR@5} & \multicolumn{1}{c|}{MRR@10} & \multicolumn{1}{c|}{HIT@3} & \multicolumn{1}{c|}{HIT@5} & \multicolumn{1}{c|}{HIT@10} & \multicolumn{1}{c|}{MRR@3} & \multicolumn{1}{c|}{MRR@5} & \multicolumn{1}{c|}{MRR@10} \\ \hline
\multicolumn{1}{|c|}{Full Model}                   & \textbf{30.56}             & \textbf{36.63}             & 44.80                       & \textbf{23.44}             & \textbf{24.83}             & \textbf{25.92}              & \textbf{69.32}             & 79.23                      & \textbf{87.23}              & \textbf{53.14}             & \textbf{55.42}             & \textbf{56.52}              \\ \hline
\multicolumn{1}{|c|}{w/o Contrastive Loss}         & 30.10                      & 36.37                      & \textbf{45.03}              & 22.96                      & 24.39                      & 25.55                       & 69.25                      & 79.18                      & 87.06                       & 53.11                      & 55.40                      & 56.49                       \\ \hline
\multicolumn{1}{|c|}{w/o UserSessionSimNet}        & 30.33                      & 36.31                      & 44.26                       & 23.29                      & 24.66                      & 25.72                       & 69.19                      & \textbf{79.42}             & 87.11                       & 53.02                      & 55.37                      & 56.44                       \\ \hline
\multicolumn{1}{|c|}{w/o Global Session Embedding} & 29.26                      & 34.93                      & 42.32                       & 22.67                      & 23.97                      & 24.95                       & 67.93                      & 78.11                      & 85.51                       & 52.42                      & 54.78                      & 55.79                       \\ \hline
\multicolumn{1}{|c|}{w/o Local Session Embedding}  & 28.69                      & 34.28                      & 41.88                       & 22.02                      & 23.30                      & 24.31                       & 65.56                      & 76.52                      & 84.51                       & 50.38                      & 52.90                      & 54.00                       \\ \hline
\end{tabular}}
\end{table*}

%% file: 060conclusions.tex
In this paper, we propose USP-GNN, an effective and novel personalized session based recommender model. With additional contrastive loss and user-session attention module, proposed model has exceeded some of the state-of-the-art performances through experimental results. We propose a view that learning of global heterogeneous graphs can be treated as an augmentation for user embedding, and can lead to better recommendation with additional contrastive loss. In future studies, developing an appropriate combining method for user embedding and session embedding will be crucial to create more effective SRS models.